\documentstyle[preprint,aps]{revtex}
\tightenlines
\begin{document}
\title{
From favorable atomic configurations to supershell structures: a new
interpretation of conductance histograms
}
\draft
\author{A. Hasmy$^1$, E. Medina$^1$ and P.A. Serena$^{1,2}$}
\address{$^1$Laboratorio de F\'\i sica Estad\'\i stica de Sistemas 
Desordenados, Centro de F\'{\i}sica, IVIC,
Apartado 21827,\\ Caracas 1020A, Venezuela}
\address{$^2$Instituto de Ciencias de Materiales de Madrid,
Consejo Superior de Investigaciones Cient\'\i ficas,\\
Cantoblanco, 28049-Madrid, Spain}
\date{Sent for publication: August 14th 2000}
\maketitle

\begin{abstract}
Simulated minimum cross-section histograms of breaking Aluminum 
nanocontacts are produced using molecular dynamics. The results allow
a new interpretation of the controverted conductance histogram peaks
based on preferential geometrical arrangments of nanocontact necks. As
temperature increases, lower conductance peaks decrease in favor of
broader and higher conductance structures. This reveals the existence of
supershell structures favored by the increased mobility of Al atoms.
\end{abstract}



Electron transport in metallic nanowires having cross-sections formed
by few atoms exhibit a rich quantum phenomenology since their
characteristic sizes are of the order of the electron Fermi wavelength
($\lambda _{\text{F}}$). Transverse confinement of electrons in the
nanocontact region only allows few propagating modes $N$, giving rise to
quantization of the conductance (assuming perfect transmittance per
channel) $G=N\times G_0$, where $G_0=2e^2/h$ is the conductance
quantum\cite{Land89}. Although conductance quantization (CQ) was
first seen in 2DEG devices\cite{2DEG}, the study of this phenomenon in
metallic nanocontacts has received much attention because of its
technological implications in mesoscopic swithching elements and new
nanoelectronics devices\cite{Nanowires97}.

Metallic nanocontacts have been built with scanning tunneling
microscopy (STM)\cite{Pascual93,Olesen94}, with 
the Mechanically Controllable Break Junction (MCBJ) method\cite
{Muller92,Krans93} as well with two plain macroscopic
wires\cite{Costa95}.  Nanocontact formation and later
breaking are usually carried out in a controlled way using piezoelectric
actuators.  The conductance evolution is followed by applying a bias
voltage to the nanocontact. In general, conductance jumps appear
during the nanowire breaking process and correspond to changes of the
transmission probability of the propagating modes or sudden variations
of $ N$. 

Since every single contact breaking experiment has its own conductance
evolution, the study of CQ has been usually addressed from a
statistical point of view, elaborating conductance histograms (CH) to
get information on the electrical behavior of nanocontacts. In
general, CH show a well-defined peaked structure, with higher
conductance probabilities close to integer values of $ G_0$. This
trend has been usually interpreted as the evidence of the CQ in
different metallic species\cite{Olesen94,Costa95,Costa97c}. CH have
been used to evaluate changes in transport properties when modifying
external parameters such as the magnetic field\cite{Costa97c}, 
the chemical environment\cite{Li98},
temperature\cite{Sirvent96,Yanson99}, 
or the applied bias
voltage\cite{Yasuda97}.

Although CH has become a standard tool to study the electronic
transport in metallic nanocontacts, there is no solid theoretical
background justifying their use. Any realistic interpretation of CH
should take into account that there is a strong coupling between
electronic and mechanical properties\cite {Rubio96}, since conductance
and force jumps are correlated, through crystalline rearrangements
inside the nanocontact as predicted by Molecular Dynamics (MD)
simulations\cite{Landman90}. Therefore, CH give information of the
conductance, while the measured histogram weights are related to the
stability of a given conductance situation. On the other hand, while
for monovalent species an interpretation of CH in terms of number of
atoms at the neck seems straightforward\cite{Oni98}, for species with higher
chemical valence the scenario is less clear.  Scheer et{\it \
al.}\cite{Scheer97} showed that three modes contribute to electron
transport in aluminum monoatomic contacts whose conductance is close
to $G_0$. Similar results for Pb and Nb demonstrate the existence of a
correspondence between valence and the number of conduction channels
participating in one-atom contacts\cite{Scheer97,Ludoph00a}. For Al,
CH constructed from many MCBJ experiments\cite {Yanson97} clearly show
peaks close to $G/G_0\sim $0.8, $\sim $ 1.8, $\sim $3.0, and $\sim
$4.0 at $T$=4K. A quite similar experimental series of peaks for Al has been
reported recently
\cite{Lud20}. These results have been used to reconsider the role of
CH to indicate CQ features. In particular, it has been suggested that
peaks in aluminum CH originate in favorable atomic configurations
appearing during nanocontact stretching\cite {Yanson97,Lud20}.

In the past, within the framework of the free-electron model together
with conductance calculations, theoretical CH have been obtained by
considering a  simplified {\em ad-hoc } nanowire dynamics
\cite{Torres96}. A realistic description of the dynamics of the
metallic nanowires breakage can be obtained by means of first-principles
calculations\cite{Nakamura99}. Other authors used classical Molecular
Dynamics accompanied with conductance calculations
\cite{Todorov93,Mehrez97,Sorensen98}. However, to construct CH, both
approaches are impracticable because of the prohibitive
computational cost.

In this letter, we have adopted a different perspective.  Our
goal is to construct {\it atomic configuration histograms} of the neck
instead of conductance histograms. In this way we will be able to know
whether there are favorable atomic configurations behind the peaked
structure obtained in CH. We have studied the atomic configuration
histograms of aluminum, a trivalent metal, which shows a well defined
and reproducible peaked structure\cite{Yanson97,Lud20}. 
For the statistical study of breaking nanocontacts we 
performed MD simulations using the embedded atom method (EAM)
\cite{Foiles86} as in previous works \cite{Landman90,Mehrez97}. In particular, we have used
state-of-the-art many-body EAM interatomic potentials for aluminum able
to fit bulk and surface properties\cite{Mishin99}. Simulations were
carried out in a wide range of temperatures (4K-450K) in order to
assess the influence of temperature $T$ on the configuration histograms. 
The temperature is controlled during the simulation using conventional
scaling of velocities.

As starting point of our MD simulation we consider a parallelepiped
supercell containing 1008 Al atoms ordered following a fcc
crystallographic structure.  The lattice constant is taken to be equal
to the experimentally measured value (4.05 \AA ). 
Aluminum atoms were initially distributed in 18 layers
perpendicular to the (111) direction containing 56 atoms each [see
Fig.1a (left) for illustration]. The
direction (111) ($z$ axis in our simulation) corresponds to that in
which the contact will be elongated until breaking, although we also
have considered other orientations discussed below. Each single MD
simulation is carried out following three stages. First, we relax the
initial bulk-like ordered system for 2000 iterations, imposing periodic
boundary conditions (PBC) in the $x$ and $y$ directions of the
sample. The time interval per iteration step in all our simulations is
$ dt=10^{-14}$ sec. At the end of this equilibration step the system
undergoes a relaxation characterized by a contraction along the $z$
axis. In the final thermally equilibrated structure two bi-layer slabs
are defined at the top and bottom of the supercell. All atomic
positions inside these slabs are frozen during subsequent MD stages,
defining the bulk support of the nanocontact during the
breaking process. In a second stage, the system is again thermally
equilibrated during 3000 iterations, but now PBC are preserved only in
the $x$ and $y$ directions of the frozen slabs. The other
atoms, defining a free nanowire evolve to reach a new equilibrium
configuration.  For temperatures higher than 450K, we found that
we are not able to stabilize the nanowire geometry, which
spontaneously breaks.  This is consistent with the decrease of the
melting temperature noted for metallic nanowires as their radius
decreases\cite{Gulseren95}. The third stage corresponds to the nanowire
stretching, separating both supporting frozen slabs at a constant
velocity of $2\times 10^{-4}$ \AA\ per iteration step. The full
determination of atomic positions during the elongation process allows
the determination of an effective surface for the nanocontact and a
minimum cross-section $S_N$ time evolution\cite{Sorensen98}. The
quantity $S_N$ is computed  as in ref. \cite{Sorensen98} (in units
of numbers of atoms) and is approximately equal to the Sharvin
conductance of the nanocontact in units of $G_0$. In order to reduce the
computational effort, we evaluate $S_N$ each 10 MD iterations.

An illustration of the nanocontact evolution during elongation $\Delta d$
is depicted in Fig. 1a. Fig. 1b shows the evolution of $S_N$ as a
function of $\Delta d$ for four different realizations (i.e. each
relaxed pallelepiped gives rise to a different nanowire evolution until
breaking takes place). The jumps in $S_N$ curves are correlated with the
jumps of the resulting force on the frozen slabs as observed
previously\cite{Pascual93,Sorensen98}. By accumulating conductance
traces we can construct minimum cross-section histograms
$H(S_N)$. The number of single nanowire breaking realizations $N_s$
considered is such that the obtained histogram peak structures are
unchanged under further averaging (see Fig. 1c). In our calculations we
have found that
$N_s=$100 provides a reliable $H(S_N)$ curve.

In Fig. 2 we plot the histogram $H(S_N)$ for three different
temperatures.  For low temperatures ($T=4$K) it is clear that
$H(S_N)$ presents a well-defined peaked structure with maxima located
at $S_N=1,2,3,4$, etc.  Therefore, {\em the statistics of the
nanocontact at the narrowest section exposes favorable atomic
configurations during its stretching}. Although our histograms do not
contain direct information on conductance, the resemblance with
experimental histograms\cite{Yanson97,Lud20} is remarkable. Assuming that 
one Al atom
provides a conductance close to $G_0$ (including components from three
different channels) the interpretation of $H(S_N)$ as a conductance
histogram is close to the observed results.

For increasing temperatures, we found that the histogram dramatically
changes loosing its peaked features for low $S_N$ values, whereas a
different structure appears for high $S_N$ values. This trend is similar
to that found in Na conductance histograms reflecting the
appearance of supershell high stability structures\cite{Yanson99}. In
our case, the temperature needed to find such supershell configurations
is higher than in Na wires due the higher cohesion energy of Al (1.13 and
3.36 eV in Na and Al bulk, respectively). Although the explored region
only corresponds to $S_N \leq 30$ the presence of peaks at $S_N \simeq
5, 8, 12, 17, 24$ is evident.  The shell structure becomes observable at
higher temperatures because of the increased mobility of the atoms. This
allows the system to explore many atomic configurations in order to
efficiently find a suitable local energy minimum.

By inspecting Fig. 2 (where $H(S_N)$ is shown in a range
similar to that used in experimental situations), it is clear that
temperature has a dramatic effect on the favorable geometrical
configurations. The higher the temperature the larger the background
structure, giving rise to a histogram with a less defined peak
structure. In particular it is very striking that the peak at $S_N=1$ is
still well defined at $T=300$K while the peak $ S_N=2$ decreases its
relative weight with respect those found in $S_N=3$ and 4, in comparison
to the 4K case. At
$T$=450K the $S_N=1$ peak almost vanishes and the broad structure
around $S_N=5$ appears, whereas intermediate peaks at $S_N=2,3$ and 4
disappear. Although these predictions correspond to cross-section
histograms, we believe a similar behavior is to be expected in true CH.

In actual experiments the nanowire orientation is not well defined,
and the histogram would contain components coming from different
nanowire orientations. In order to analyze these effects we have
studied the histograms $H(S_N)$ at $T=4$K for two other orientations
of the stretching direction: (011) and (001). These results are
depicted in Fig. 3. It is worth noticing that the histograms have
now a more complicated structure but also reflect the presence of
favorable atomic configurations.  We can then expect that in 
experiments the histograms can be understood in terms of a combination of
the histograms appearing in Fig. 3. However it is clear that a
structure with peaks at $S_N=1$ and 2 should be present. For $ S_N\geq
3$ the peaked structure should become less clear and would depend on the
experimental conditions (as it happens when comparing two different
experimental conductance histograms in
Al\cite{Yanson97,Lud20}).  Another relevant observation is that for less
stable configurations, as those considered when stretching along (011)
and (001) directions, the shell structure tends to appear at lower
temperatures [see, for instance, the peak at $S_N=5$ formed in the (011)
case]. Thus, orientation effects will also contribute in bringing out
supershell effects. 

In conclusion, we have obtained, for the first time,
minimum cross-section histograms for Al nanowires breaking MD
simulations using state-of-the-art EAM potentials to describe Al-Al
interactions. At low temperatures we found that there exist favorable
atomic arrangements as suggested some time ago by 
experimentalist\cite{Yanson97,Lud20}.
This provides a new way to interpret experimental conductance histograms
on the basis of MD simulations. For increasing temperatures, the system
is able to explore more atomic configurations where low conductance
regions show less structure and appearing well defined high stability
regions for higher conductance values appear, as reported for
Na\cite{Yanson99}. In the low conductance region it is found that at
$T=300$K the histogram peak at
$S_N=2$ decreases its relative weight with respect the peak $S_N=3$.
That is, at 4K the peak height sequence is 1, 2, 3, 4 whereas at 300K
the sequence change to 1, 4, 3, 2. For $T$=450K the peak at $S_N=1$
almost vanishes, and the presence of supershell structure becomes
evident. Finally, for other nanowire stretching directions in the low
temperature regime we also found that there exist favorable
configurations, although the histogram presents a more complex
structure. However peaks at $S_N=1$ and 2 remain. Nanowires with lower
stability tend to form the supershell structure at lower temperatures
than those with high initial stability. 

We acknowledge J.J. S\'aenz, P. Garc\'\i a-Mochales, C. Urbina, R.
Paredes, J.R. Villarroel, and A. Garc\'\i a-Mart\'\i n, for helpful
discussions. This work has been partially supported by the CSIC-IVIC 
researchers exchange program and the Spanish DGICyT (MEC) through
Project PB98-0464.

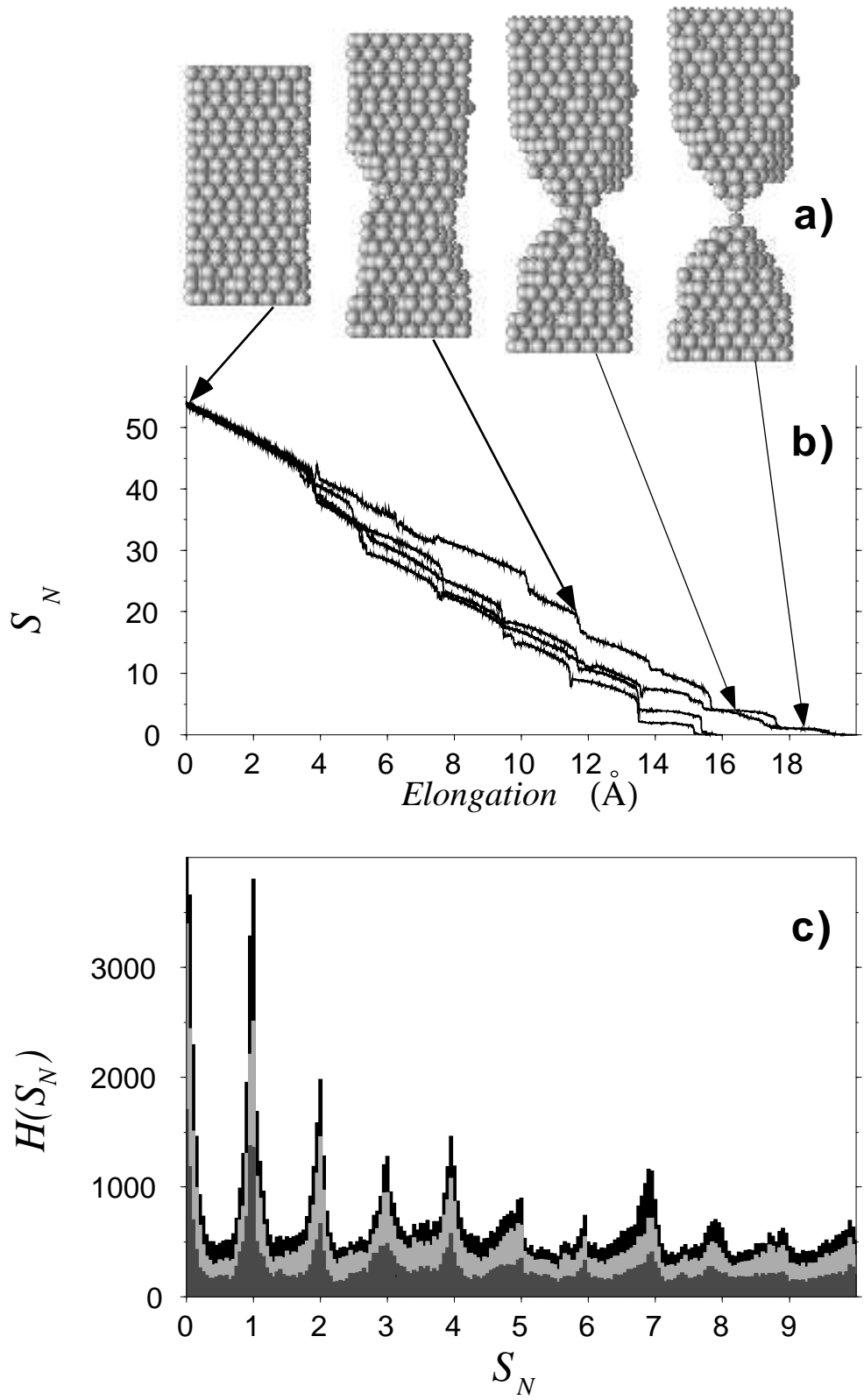
\begin{figure}[tbp]
\caption{(a) Typical nanocontact configurations during the elongation of 
the sample. (b) Minimum 
cross-section $S_N$ evolution for
four different Al nanowires at $T=4$K. The stretching direction is 
parallel
to the (111) direction, and the arrows indicate the corresponding $S_N$ 
for
each configuration. 
(c) Minimum cross-section histograms $H(S_N)$. Dark gray, gray and black
curves correspond to histograms obtained with different number of
realizations (35, 70 and 100, respectively). }
\label{fig1}
\end{figure}

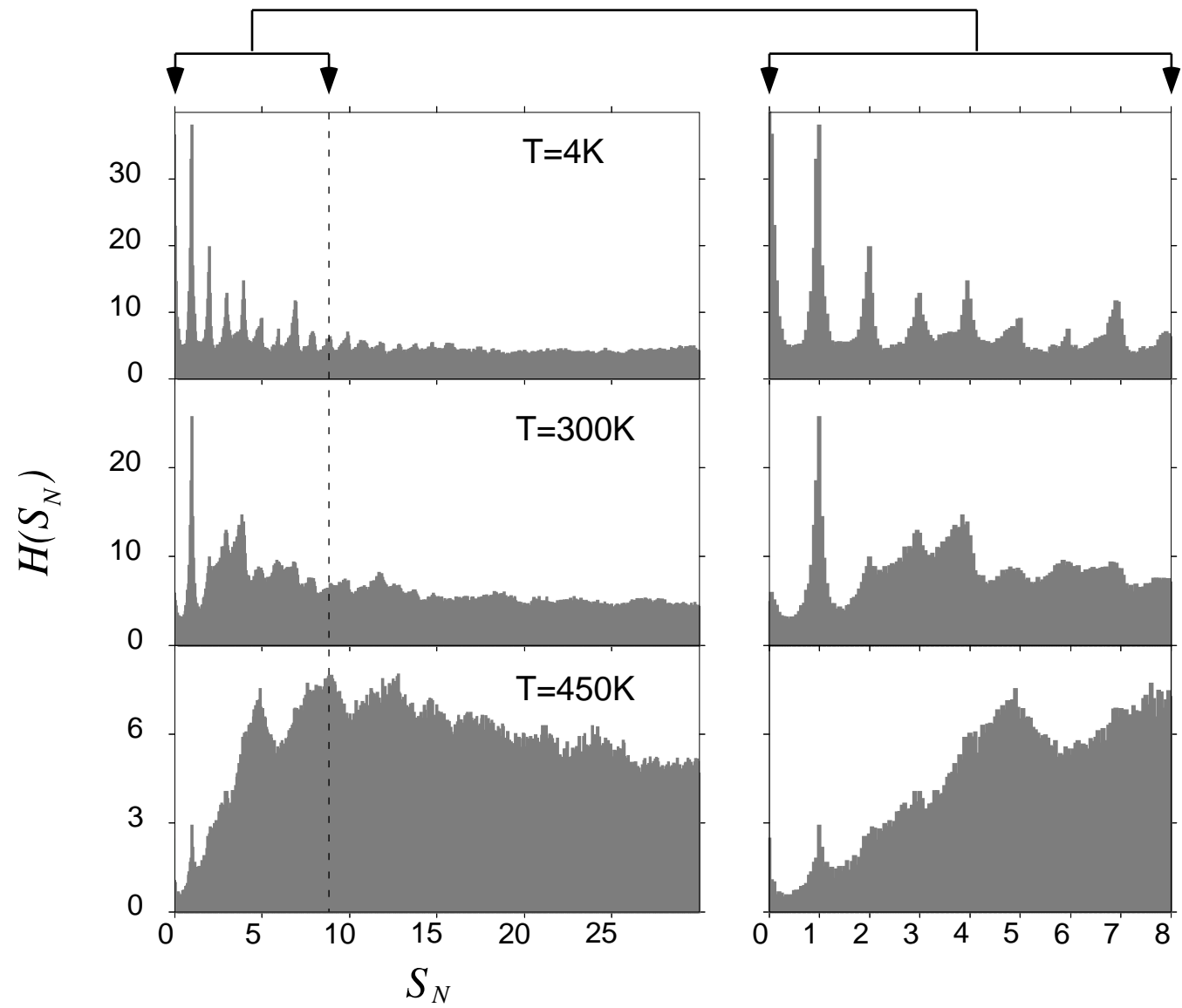
\begin{figure}[tbp]
\caption{ Minimum cross-section histograms $H(S_N)$ for stretching of
Al nanowires along the (111) direction at $T$=4K, 300K, and 450K.
The right panel is a zoom of the range $0 \leq
S_N \leq 8$. The curves are normalized to the number of single breaking
events (100 per
temperature value).}
\label{fig2}
\end{figure}

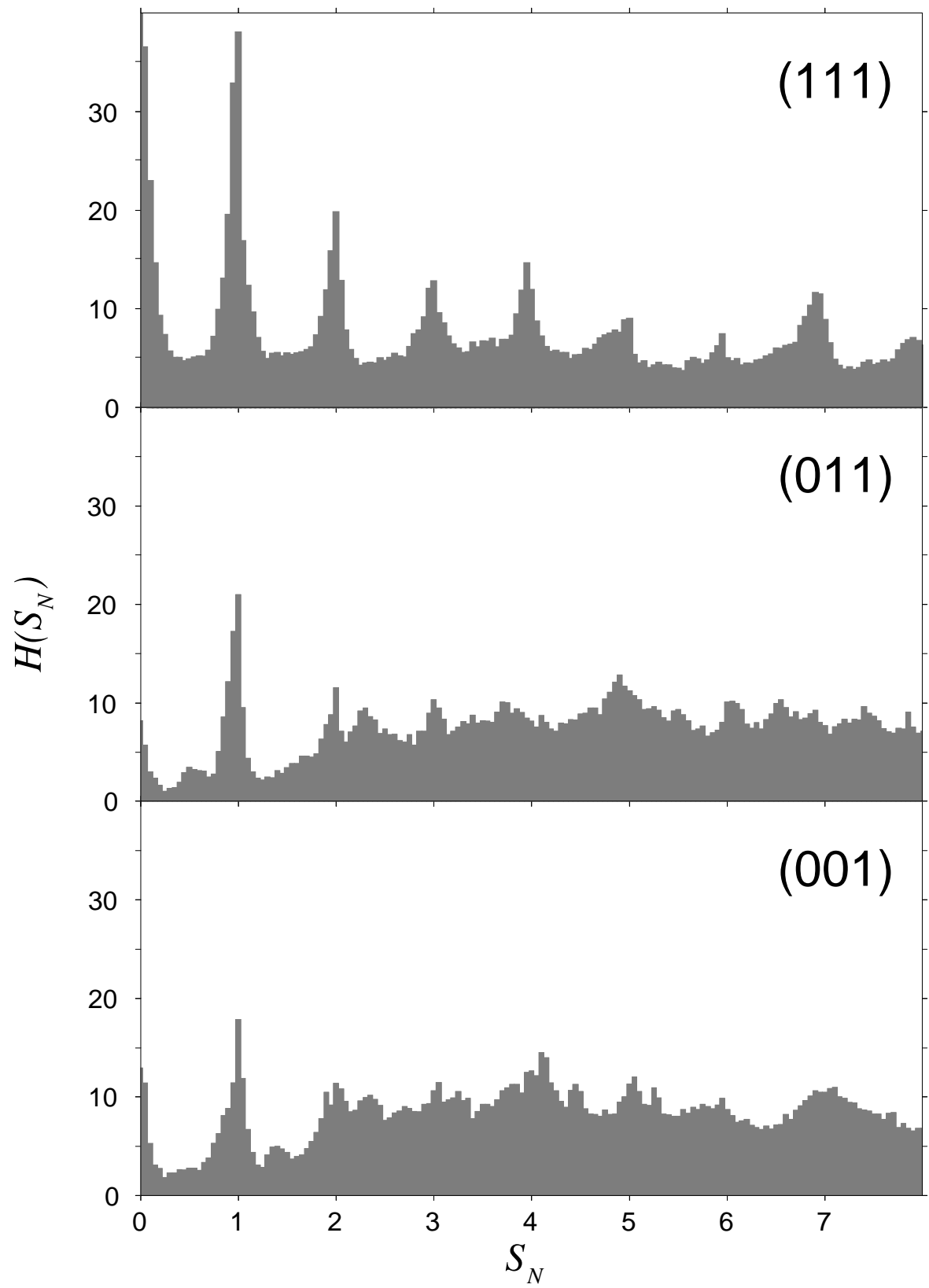
\begin{figure}[tbp]
\caption{ Comparison of the 
minimum cross-section histograms $H(S_N)$ of aluminum nanowires at 
$T=4$K stretching along three different orientations (111), (011) and
(001). The last two histograms are constructed with 70 breaking
events and nanocontacts contain 1000 and 980 atoms, respectively. All
histograms are normalized to the number of realizations}
\label{fig3}
\end{figure}


\end{document}